\documentclass[a4paper, amsfonts, amssymb, amsmath, apssamp, nofootinbib, reprint, prl, aps, showkeys, twoside,superscriptaddress,noeprint]{revtex4-2}

\usepackage[english]{babel}

\usepackage{physics}

\usepackage{graphicx}% Include figure files
\usepackage{dcolumn}% Align table columns on decimal point
\usepackage{bm}% bold math
\usepackage{xcolor}
\usepackage{ulem}
\usepackage{comment}
\usepackage{appendix}
\begin{document}

\preprint{APS/123-QED}

\title{One-dimensional phase singularities as a source of atypical local oscillations}

\author{Enrique G. Neyra}
\email{enrique.neyra@ib.edu.ar}
   \affiliation{Instituto Balseiro (Universidad Nacional de Cuyo and Comisión Nacional de Energía Atómica) and CONICET CCT Patagonia Norte. Av. Bustillo 9500, Bariloche 8400 (RN), Argentina. }%\\This line break forced with \textbackslash\textbackslash

\author{Laureano A. Bulus Rossini}
   \affiliation{Instituto Balseiro (Universidad Nacional de Cuyo and Comisión Nacional de Energía Atómica) and CONICET CCT Patagonia Norte. Av. Bustillo 9500, Bariloche 8400 (RN), Argentina. }

\author{Fabi\'an Videla}
\affiliation{Centro de Investigaciones \'Opticas (CICBA-CONICET-UNLP), Cno.~Parque Centenario y 506, P.O. Box 3, 1897 Gonnet, Argentina}
\affiliation{Departamento de Ciencias B\'asicas, Facultad de Ingenier\'ia UNLP, 1 y 47 La Plata,Argentina}%\\This line break forced with \textbackslash\textbackslash

\author{Pablo A. Costanzo Caso}
   \affiliation{Instituto Balseiro (Universidad Nacional de Cuyo and Comisión Nacional de Energía Atómica) and CONICET CCT Patagonia Norte. Av. Bustillo 9500, Bariloche 8400 (RN), Argentina. }
   
\author{Lorena Reb\'on}
\email{rebon@fisica.unlp.edu.ar}
\affiliation{ 
Departamento de F\'isica, FCE, Universidad Nacional de La Plata, C.C. 67, 1900 La Plata, Argentina%\\This line break forced with \textbackslash\textbackslash
}%
\affiliation{Instituto de F\'isica de La Plata, UNLP - CONICET, Argentina}

\date{\today}% It is always \today, today,

\begin{abstract}

In this work, phase singularities embedded in a wavepacket are shown to act as sources of atypical localized oscillations when the packet interacts with a linear system. We refer to these oscillations as \textit{chirpons}, since they arise as strong variations of the instantaneous frequency (chirp). A mathematical expression is then provided to describe \textit{chirpons}, and their behavior is explored through the interaction of a super-bandwidth wavepacket—containing two singularities—with a damped harmonic oscillator, a fundamental model for many physical systems. This interaction is analyzed theoretically, and the predictions are verified experimentally using a resonant electrical circuit as a realization of the oscillator. The results show that \textit{chirpons} evolve in a manner fundamentally different from standard Fourier oscillations, revealing features of linear systems that are otherwise inaccessible. This introduces a new approach to analyze and characterize system responses, with potential applications in high-resolution spectroscopy and signal sensing.
\end{abstract}

\maketitle

%{\textit{Introduction}}.---
In its complex form, a temporal wavepacket (WP) can be described mathematically as $f_w(t)=A_w(t)e^{i\Phi_w(t)}$, where $A_w(t)$ and $\Phi_w(t)$ are real functions representing the envelope and the temporal phase, respectively. From this representation, the instantaneous frequency $\omega_w(t)=\frac{d\Phi_w(t)}{dt}$ determines the local oscillations of the WP. While in most cases $\omega_w(t)$ remains within the Fourier spectrum of $f_w(t)$, for certain specially structured WPs this is not the case, giving rise to superoscillatory and suboscillatory 
behavior whenever $\omega_w(t)$ exceeds or falls below the spectral bounds.
These effects arise from destructive interference and can occur in any wave phenomenon, always confined to finite temporal windows, which indicates that the local dynamics of a WP differ significantly from its global spectral properties. 

Superoscillatory phenomena---and, to a lesser extent, suboscillatory ones---were first examined in a purely mathematical context~\cite{berry2006evolution,aharonov2017mathematics}, eventually finding applications in subdiffractive
beams~\cite{zheludev2022optical,cheng2022super,wu2019broadband,zhang2023generating}, signal processing~\cite{ferreira2006superoscillations,luo2023creation}, ultrashort pulses~\cite{eliezer2017breaking}, and acoustic waves~\cite{Brehm2020,shen2019ultrasonic}, among others~\cite{chen2019superoscillation,lin2023reconfigurable,dennis2008superoscillation,yuan2016quantum,eliezer2014super}.
On the other hand, phase singularities in wave fields---points of zero amplitude where the phase is undefined---constitute another distinctive manifestation of interference~\cite{nye1974dislocations,freund1994wave,dennis2001topological}. Indeed, singular optics~\cite{Soskin2001} explores a wide range of effects directly related to singularities in light~\cite{berry2000making,berry2023singularities}, which exhibit features absent in smooth wavefronts. Thus, within this framework, super-- and suboscillations can be naturally associated with phase singularities, since both phenomena originate from interference effects~\cite{berry2008natural}.

In Ref.~\cite{Neyra21}, we presented the superbandwidth (SB) phenomenon in laser pulses, showing that the SB pulse interacts locally with matter as if it possessed an effective bandwidth that exceeds its Fourier spectrum. Later, in Ref.~\cite{neyra2024superbandwidth}, we studied the propagation of these pulses in a dispersive medium characterized by a quadratic spectral phase. Chromatic dispersion induces a variation in the instantaneous frequency (chirp), and as the SB pulse propagates, two local oscillations emerge, one oscillating faster and the other slower than the highest and lowest frequencies of its Fourier spectrum, respectively. This behavior was further analyzed in the context of the free propagation of a non-relativistic quantum particle, whose initial state is an SB--WP in momentum--position~\cite{neyra2025free}.

Against this background, while phase singularities and superoscillatory behavior are usually discussed as structural properties of the wave field, many realistic platforms probe the system response rather than the bare input field. This naturally raises a largely unexplored question: what observable dynamical signatures do temporal phase singularities generate after interacting with a linear system?
  
Building on our earlier results, in this work we demonstrate that the emerging oscillations---here referred to as \textit{chirpons}---originate from temporal phase singularities. The term ``chirpon" is introduced to emphasize the time-varying nature of such oscillations, analogous to that of chirped signals. We then analyze the interaction between a SB signal and various linear systems---dispersive propagation and resonant interactions as typical scenarios---both theoretically and experimentally, revealing how \textit{chirpons} evolve and how their behavior differs from that of conventional Fourier frequencies. Furthermore, we show that a local topological charge can be assigned to each phase singularity, whose absolute value constitutes a conserved quantity in the interaction of the associated \textit{chirpon} with linear systems.

\begin{figure}[h!]
\includegraphics[width=0.5\textwidth]{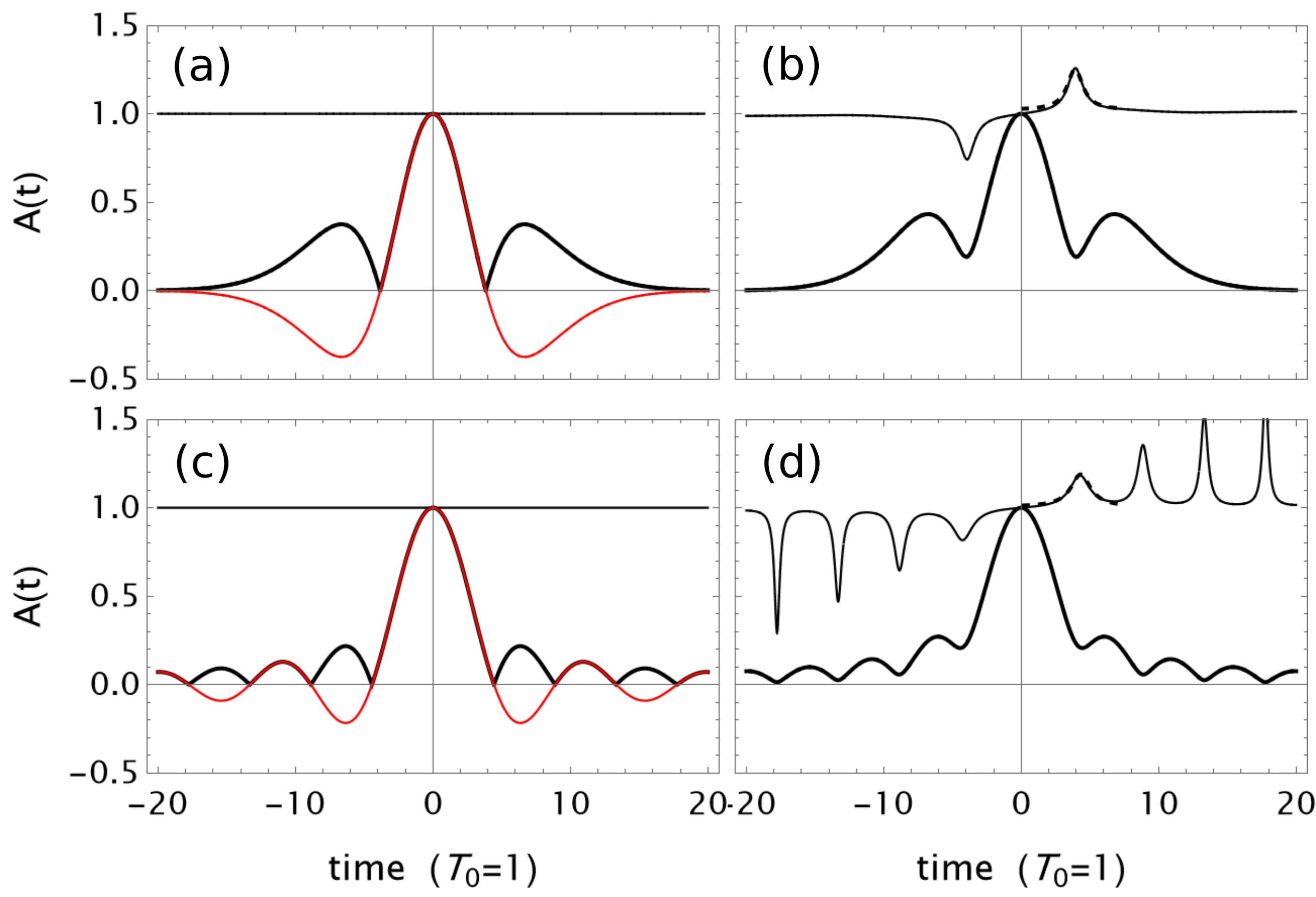}
\caption{(a) SB--WP with parameters $\alpha=1$, $\beta=0.5$, and $\Delta\omega=0.5$. The red line corresponds to the term in parentheses in Eq.\eqref{e2}, while the thick and thin black lines show the real amplitude $\left|E_{SB}(t)\right|$ and the normalized instantaneous frequency $\omega_{SB}(t)/\omega_0$, respectively. Phase singularities appear at $t=\pm t_z$ (black arrows). (b) $\left|E_{SB}(t)\right|$ and $\omega_{SB}(t)/\omega_0$ after propagation through a dispersive medium with $\Omega = 2$. The dashed line shows the fit of $\omega_C^+(t,0.69T_0)/\omega_0$ (Eq.\eqref{e9} with $\gamma = +0.69T_0$) to $\omega_{SB}(t)/\omega_0$. (c) and (d): Same analysis for a WP given by $E_{sinc}(t)=\mathrm{sinc}(t/\sqrt{2})e^{i\omega_0 t}$. In all panels, time is normalized to the carrier period $T_0=2\pi/\omega_0=1$.} \label{fig1}
\end{figure}

\textit{Formalism.}--- An SB--WP can be generated by destructive interference between two Gaussian functions with different widths, and centered at the same carrier frequency $\omega_0$~\cite{neyra2021tailoring,neyra2022simple}. 
In the spectral domain, this is described mathematically by the expression 
\begin{equation}
\tilde{E}_{SB}(\omega)=e^{-(\frac{\omega-\omega_0}{\Delta\omega})^2}-\alpha e^{-(\frac{\omega-\omega_0}{\beta\Delta\omega})^2},\label{e1}
\end{equation}
while its temporal description is obtained  
by Fourier transforming back Eq.~\eqref{e1}: 
\begin{equation}
E_{SB}(t)=\left(e^{-\frac{(\Delta\omega t)^2}{4}}-  
\alpha\beta e^{-\frac{(\beta\Delta\omega t)^2}{4}}\right)e^{i\omega_0 t}.\label{e2}
\end{equation}
In Fig.~\ref{fig1}(a), the SB--WP corresponding to the synthesis parameters $\alpha = 1$, $\beta = 0.5$ and $\Delta\omega = 0.5$, is shown in
the time domain. The 
red line depicts the term in parentheses in Eq.~\eqref{e2}. 
From this plot, it is clear that $E_{SB}(t)$
can be written in polar form $E_{SB}(t) = \left|E_{SB}(t)\right|e^{i\Phi_{SB} (t)}$, with
\begin{eqnarray}\label{e4}
\left|E_{SB}(t)\right|&=&\left|e^{-\frac{(\Delta\omega t)^2}{4}}-  
\alpha\beta e^{-\frac{(\beta\Delta\omega t)^2}{4}}\right|,\\
\Phi_{SB} (t)&=&\omega_0 t+ \pi \Theta(t-t_z) + \pi \Theta(-t-t_z)|\nonumber
\end{eqnarray}
where $\Theta(t)$ denotes the Heaviside step function. Here, the values $t = \pm t_z$, which depend on the pulse synthesis parameters $\alpha$, $\beta$ and $\Delta\omega$, correspond to the instants at which $E_{SB}(t)$ vanishes. Therefore, $\left|E_{SB}(t)\right|$ and $\Phi_{SB} (t)$, as defined in Eq.~\eqref{e4}, %respectively, 
provide the correct real amplitude and temporal phase for the SB--WP. As a result, the instantaneous frequency of $E_{SB}(t)$ can now be readily obtained: 
\begin{equation}
\omega_{SB} (t)\equiv\frac{d\Phi_{SB} (t)}{dt}=  \omega_0 + \pi \delta(t-t_z) - \pi \delta(-t-t_z),\label{e6}
\end{equation}
where the Dirac delta functions in the expression of $\omega_{SB} (t)$, arise from the phase singularities at $t=\pm t_z$ described by the Heaviside terms $\Theta(\pm t-t_z)$. These features are also illustrated in Fig.~\ref{fig1}(a), where the thin black line represents the normalized instantaneous frequency $\omega_{SB}(t)/\omega_0$, and the Dirac delta contributions are schematically indicated as black arrows. From that, it can be seen that within the time interval $\left[-t_z, t_z\right]$, the SB--WP has a phase $\Phi_{SB} (t)-\omega_0 t=0$. Outside this region---i.e., for $t \in \left(-\infty, -t_z\right) \cup \left(t_z, \infty\right)$---the phase becomes $\Phi_{SB} (t)-\omega_0 t=\pi$. This transition corresponds to a phase jump of $\pi$ at $t = \pm t_z$.

We now consider an analytical representation of the Dirac delta function, to aid in characterizing the \textit{chirpons} as local oscillations of the SB--WP resulting from its interaction with a linear system.
The Heaviside step function describing a $\pi$--phase jump at $t = t_z$ can be expressed as $\Theta(t - t_z) = \lim_{\gamma \rightarrow 0} \left[ \frac{1}{\pi} \arctan\left( \frac{t - t_z}{\gamma} \right) + \frac{1}{2} \right]$,
and the Dirac delta function is then obtained by differentiating this expression:
\begin{equation}
\delta(t - t_z) = \frac{d\Theta(t - t_z) }{dt} =\lim_{\gamma \rightarrow 0} \frac{1}{\pi} \frac{\gamma}{(t - t_z)^2 + \gamma^2}. \label{e8}
\end{equation}

As demonstrated in Ref.~\cite{neyra2024superbandwidth}, when the SB--WP interacts with a linear system characterized by a transfer function of the form $T_D(\omega) = e^{-i\Omega(\omega - \omega_0)^2}$---i.e, a dispersive medium---, the instantaneous frequency $\omega_{SB}^D(t)=\frac{d\Phi_{SB}^D(t)}{dt}$, associated with the field $E_{SB}^D(t)\equiv \mathcal{F}^{-1}\left\{T_D(\omega)\tilde{E}_{SB}(\omega)\right\}$, behaves quite differently from what one would expect. In particular, two \textit{chirpons} emerge, one to the left and one to the right of the SB--WP maximum. 

In Fig.~\ref{fig1}(b), we show the SB--WP from Fig.~\ref{fig1}(a) after propagation through a system whose $T_D(\omega)$ corresponds to $\Omega = 2$ (thick black line), together with the normalized instantaneous frequency profile, $\omega_{SB}^D(t)/\omega_0$, shown in the upper part of the figure (thin black line). Thus, guided by the features observed in this figure and considering Eq.~\eqref{e6}, along with the representation of the Dirac delta function in Eq.~\eqref{e8}, it is natural to assume that \textit{chirpons} will admit a closed-form analytical description given by the expression 
\begin{eqnarray}\label{e9}
\omega_C^{\pm}(t,\gamma) = \omega_0 + \frac{\gamma}{(t \;\mp \;t_z)^2 + \gamma^2}\;,
\end{eqnarray}
for some finite value of $\gamma$ $(\gamma \neq 0)$. The second term in the expression of Eq.~\eqref{e9} is a Lorentzian function whose full width at half maximum (FWHM) is $2|\gamma|$. This function characterizes the \textit{chirpon} localized to the right, $\omega_C^{+}(t,\gamma)$, and to the left, $\omega_C^{-}(t,\gamma)$, of $t=0$. In fact, the \textit{chirpons} observed in Fig.~\ref{fig1}(b) are well described by  Eq.~\eqref{e9} with $\gamma = \pm 0.69T_0$ 
($T_0=2\pi/\omega_0$). In that figure, the dashed black line in the upper part, which corresponds to $\omega_C^+(t,0.69T_0)/\omega_0$, accurately captures the instantaneous frequency profile $\omega_{SB}^D(t)/\omega_0$, in the vicinity of the phase singularity at $t=+t_z$.
Note also that the peak values of $\omega_{SB}^D(t)$, occurring at $t=\pm t_z$, are predicted by Eq.~\eqref{e9}: $\mathrm{max}\{\omega_{SB}^D(t)\}\equiv \omega_{max} = \omega_C^{\pm}(t_z,\gamma)=\omega_0+\frac{1}{\gamma}$ for $\gamma > 0$, and $\mathrm{min}\{\omega_{SB}^D(t)\}\equiv \omega_{min}= \omega_C^{\pm}(t_z,\gamma)=\omega_0-\frac{1}{|\gamma|}$ if $\gamma < 0$. These peak values, $\omega_0 \pm 0.23\omega_0$, lie outside the range
of the components in the spectrum of the SB--WP.

A different WP, $E_{sinc}(t)=\mathrm{sinc}\left(\frac{t}{\sqrt{2}}\right)e^{i\omega_0 t}$, is shown in Fig.~\ref{fig1}(c). Since the sinc function vanishes at infinitely many time instants, this WP exhibits an infinite number of phase singularities. In Fig.~\ref{fig1}(d), we show 
the result of its interaction with the same dispersive medium $T_D(\omega)$ as in Fig.~\ref{fig1}(b) ($\Omega = 2$). As seen in this figure, 
a \textit{chirpon} emerges at each zero of $E_{sinc}(t)$, resulting in multiple localized events (thin black line at the top), each characterized by
%a \textit{chirpon} (thin black line at the top) emerges at each temporal position where $E_{sinc}(t)$ vanishes, each one associated with 
a distinct peak instantaneous frequency, either 
$\omega_{max}$ or $\omega_{min}$. 
The dashed black line corresponds to the curve $\omega_C^+(t,0.87T_0)/\omega_0$ in the vicinity of one of the phase singularities at $t=+t_z$, where the first \textit{chirpon} to the right of $t=0$ is located.
As in the previous example shown in Fig.~\ref{fig1}(b), the function
$\omega^+_C(t,\gamma)$ accurately captures the instantaneous frequency profile of the selected \textit{chirpon}. Although not shown here, the same level of agreement is found for the remaining ones, including the \textit{chirpons} to the left of $t=0$. 
Equivalent results, but in the spatial domain, were presented in Ref.~\cite{stern2024light}.

We then ask whether \textit{chirpons} also arise in other linear systems, and how these oscillations may induce a qualitatively different system response compared to Fourier spectral components. To address this, we analyze the response of a damped harmonic oscillator, which provides a simple mathematical framework used to model a range of physical systems.
The dynamic equation of such oscillator drived by an external force $f(t)$ can be written as
\begin{equation}
\frac{d^2 x(t)}{dt^2}+\frac{\omega_r}{Q}\frac{d x(t)}{dt}+\omega_r^2 x(t)=f(t),\label{e16}
\end{equation} 
where $\omega_r$ is the resonance frequency of the system and $Q$ is its quality factor. This system is easily solved in the frequency domain multiplying the transfer function $T_A(\omega)=\frac{\omega_r^2}{\omega^2-\omega_r^2+i\omega_r\omega/Q} $ by the Fourier transform of $f(t)$, obtaining  $\tilde{x}(\omega)=T_A(\omega)\tilde{f}(\omega)$. 
\begin{figure}[h!]
\includegraphics[width=0.5\textwidth]{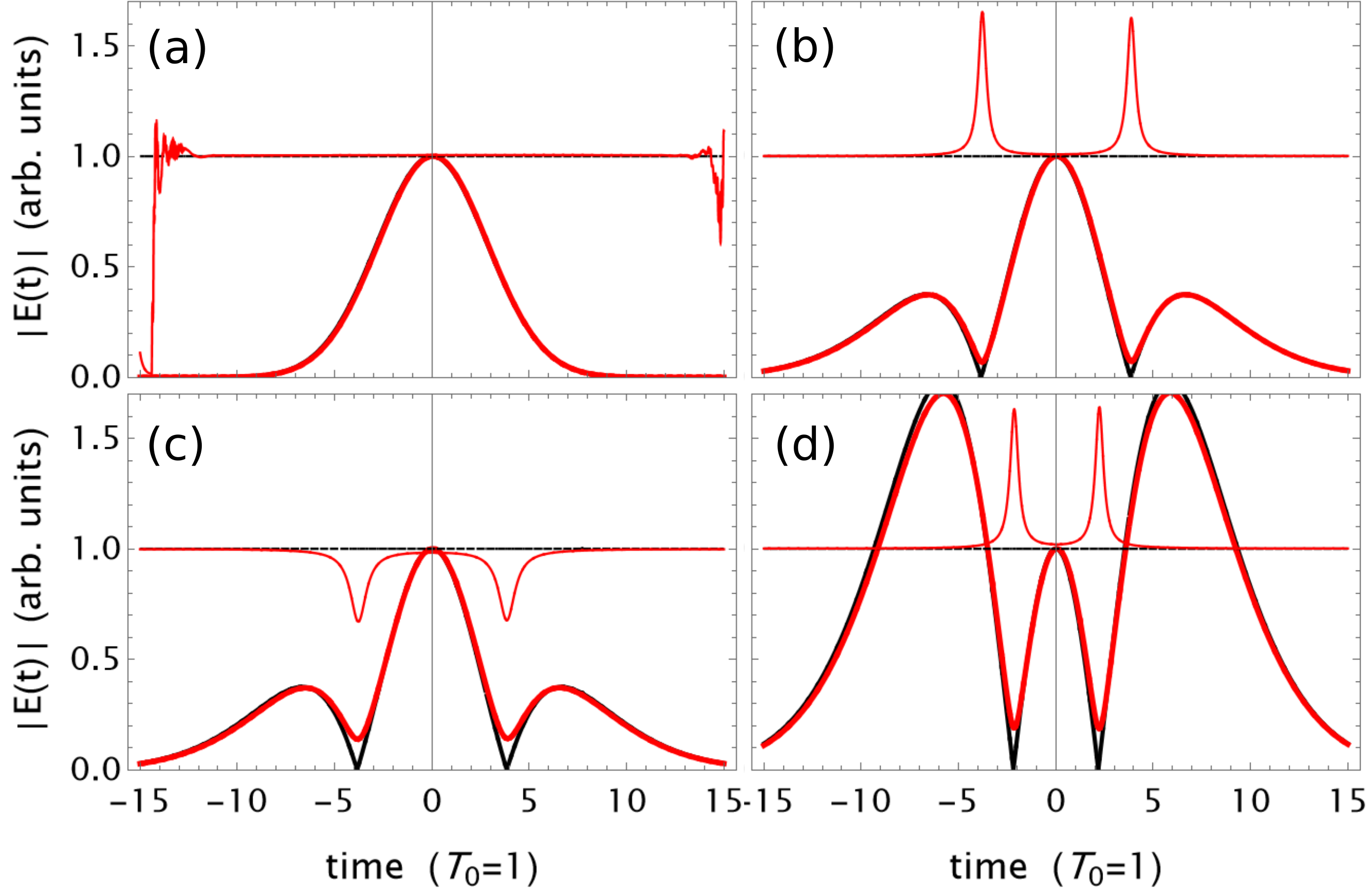}
\caption{Response (continuous red line) of a damped harmonic oscillator with quality factor $Q=10$ and varying resonance frequency $\omega_r$, to different SB input signals (continuous black line). Dashed lines show the corresponding normalized instantaneous frequency $\omega_{SB}(t)/\omega_0$: black for the input signals and red for the oscillator response.
(a) $\alpha=0$ (Gaussian input) with $\omega_r=1.5\omega_0$. (b) $\alpha=1$ with $\omega_r=1.5\omega_0$. (c) $\alpha=1$ with $\omega_r=0.6\omega_0$. (d) $\alpha=1.6$ with $\omega_r=1.5\omega_0$. %In all panels, time is normalized to the carrier period $T_0=2\pi/\omega_0=1$.
}
\label{fig2}
\end{figure}

Figure~\ref{fig2} illustrates the response of the damped harmonic oscillator to a driving force $f(t)$---also referred to as the input signal $E_{in}(t)$---which has the same functional form as the SB--WP defined in Eq.~\eqref{e2}. In all panels, the input signal is shown as a black line, and the output signal $E_{out}(t)$---i.e., the system response $x(t)$---as a red line. 
In Fig.~\ref{fig2}(a), $E_{in}(t)$ corresponds to the particular case where $\alpha = 0$, i.e., a Gaussian WP. The oscillator parameters are set to $Q = 10$ and $\omega_r = 1.5\omega_0$. It is observed here that the input and output signals are very similar, differing only by a scaling factor and a global phase, while remaining identical in shape.
Figure~\ref{fig2}(b) displays the response to the SB--WP with $\alpha = 1$, using the same system parameters than before. The output exhibits the emergence of two \textit{chirpons}, with peak frequencies around $1.65\omega_0$, which corresponds to $\gamma = 0.245T_0$.
In Fig.~\ref{fig2}(c), the system parameters are changed to $Q = 10$ and $\omega_r = 0.6\omega_0$. In this case, two \textit{chirpons} appear with peak frequencies near $0.65\omega_0$, corresponding to $\gamma = -0.45T_0$.
Finally, Fig.~\ref{fig2}(d) shows the response to a SB signal with $\alpha = 1.6$, using the same system parameters as in Figs.~\ref{fig2}(a) and \ref{fig2}(b). This case exhibits the same \textit{chirpons} as in Fig.~\ref{fig2}(b), with $\gamma = 0.245T_0$, but localized closer to $t=0$.  

From Figs.~\ref{fig2}(b), (c), and (d), it can be seen that the input signal undergoes noticeable changes around the phase singularity as it passes through the system $T_A(\omega)$, while other regions remain almost unaffected as also occurs for the input Gaussian signal shown in Fig.~\ref{fig2}(a).
Another important feature observed in Fig.~\ref{fig2} is the behavior of the \textit{chirpons} arising from the interaction with the harmonic oscillator. In all cases, the peak frequencies ($\omega_{\text{max}}$ and $\omega_{\text{min}}$) are found to be close to the system's resonance frequency $\omega_r$, regardless of whether $\omega_r > \omega_0$ or $\omega_r < \omega_0$.
Moreover, the \textit{chirpons} shown in Figs.~\ref{fig2}(b) ($\alpha = 1$) and \ref{fig2}(d) ($\alpha = 1.6$) are identical. This behavior contrasts with the response of the system $T_D(\omega)$, where the \textit{chirpons} $\omega^{\pm}_C(t,\gamma)$ swap their positions as $\alpha$ is varied, as was discussed in Ref.~\cite{neyra2024superbandwidth}.

The above results raise an open question: 
Given a linear system $T(\omega)$, what are the characteristics of the \textit{chirpons} emerging from its interaction with an SB signal? In particular, how does the \textit{chirpon} parameter $\gamma$ depends on the properties of $T(\omega)$?
\begin{figure}[h!]
\includegraphics[width=0.5\textwidth]{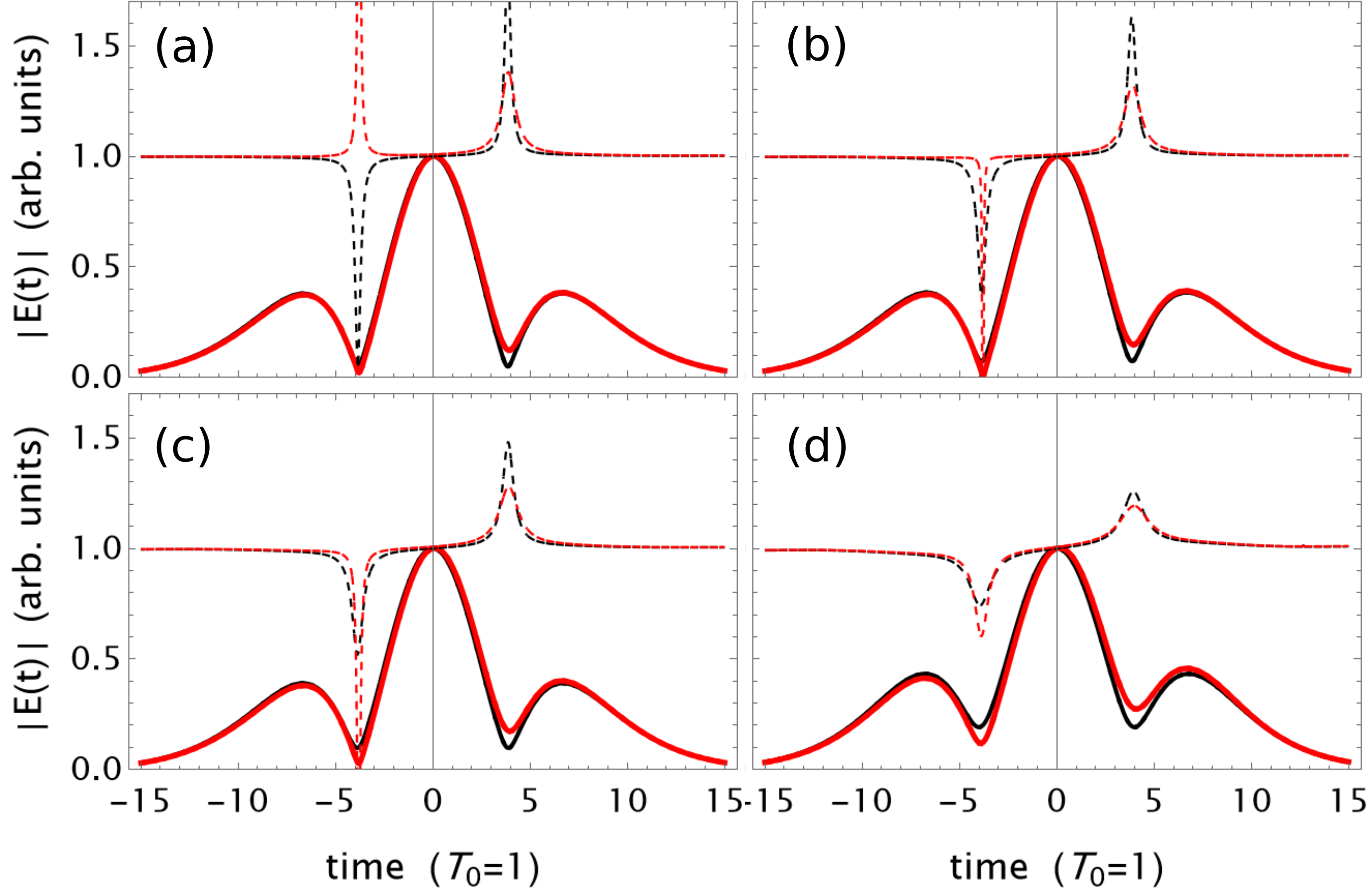}
\caption{
Response (continuous red line) of a damped harmonic oscillator to different SB input signals (continuous black line) that have previously propagated through a dispersive medium characterized by $\Omega$. Dashed lines show the corresponding normalized instantaneous frequencies $\omega_{SB}(t)/\omega_0$: black for the input signals and red for the oscillator response. The oscillator parameters are fixed at $Q=10$ and $\omega_r=1.5\omega_0$. Panels (a)–(d) correspond to $\Omega=0.5$, 0.75, 1, and 2, respectively.
} \label{fig3}
\end{figure}
With the aim to address this question, we introduce a compact notation for the \textit{chirpons} arising in SB--WPs and for the linear systems with which they interact. The state of the \textit{chirpons} is denoted as $C^{\gamma_+}_{\gamma-}$, where $\gamma_-$ and $\gamma_+$ specify the left and right \textit{chirpon}, respectively (see Eq.~\eqref{e9}). 
In this notation, $\gamma_{\pm}$ are expressed in temporal units of $T_0=1$. In the case of a ``pure" SB--WP, i.e., when $\gamma_{\pm} \rightarrow 0$, the notation is reduced to $C$. Accordingly, the \textit{chirpons} in Fig.~\ref{fig1}(b) are written as $C^{+0.69}_{-0.69}$, while those in Fig.~\ref{fig2} correspond to $C^{+0.245}_{+0.245}$ (Fig.~\ref{fig2}(b)), $C^{-0.45}_{-0.45}$ (Fig.~\ref{fig2}(c)), and $C^{+0.245}_{-0.245}$ (Fig.~\ref{fig2}(d)).

Within this formalism, we denote a linear system $T(\omega)$ by the symbol $H$. By comparing the SB--WP before and after interacting with such a system, as shown in Fig.~\ref{fig1} for the dispersive medium $T_D(\omega)$ and Fig.~\ref{fig2} for the harmonic oscillator $T_A(\omega)$, the transformation can be consistently expressed as
\begin{eqnarray}\label{eq:NewNotation}
C \otimes H^{\gamma_+}_{\gamma_-} = C^{\gamma_+}_{\gamma_-},
\end{eqnarray}
where “$\otimes$” refers to the interaction between the SB--WP $C$ and the linear system $H^{\gamma_+}_{\gamma_-}$. This notation describes how the input state $C$ is mapped into the pair of \textit{chirpons} $C^{\gamma_+}_{\gamma_-}$, so that the system itself can be regarded as \textit{hosting its own chirpons}, with $\gamma_-=-\gamma_+$ for the dispersive case ($\gamma_+ > 0$ if $\Omega > 0$, $\gamma_+ < 0$  if  $\Omega < 0$) and $\gamma_-=\gamma_+$ for the harmonic case ($\gamma_+ > 0$  if $\omega_r>\omega_0$, $\gamma_+ < 0$  if  $\omega_r<\omega_0$).  
 
 We now analyze the interaction between a pair of \textit{chirpons} originating from a dispersive medium with $\Omega>0$, and a harmonic oscillator characterized by $\gamma'$ ($\gamma'_+ = \gamma'>0$). %This originates the following interaction: $C^{+\gamma}_{-\gamma}\otimes H^{\gamma_+^h}_{\gamma_-^h} = C^{S_lS_r}_{\gamma_l\gamma_r}$. 
 The oscillator parameters are fixed at $Q=10$ and $\omega_r=1.5\omega_0$, which imply that $\gamma'=0.245T_0$ and $\omega_{max}=1.65\omega_0$. %We then 
 Figure~\ref{fig3} illustrates this interaction, $C^{\gamma_+}_{\gamma_-}\otimes H^{\gamma'_+}_{\gamma'_-}=C^{\gamma''_+}_{\gamma''_-}$\;, where four different input states $C^{\gamma_+}_{\gamma_-}$ ($\gamma_+=\gamma>0$) are considered by varying the dispersive parameter $\Omega$. In each panel, the black and red lines represent the input and output signal after interaction with the harmonic oscillator, respectively; continuous lines indicate the envelope, while dashed lines show the normalized instantaneous frequency. Panels (a)–(d) correspond to $\Omega = 0.5$, $0.75$, $1$, and $2$, with input states $C^{+0.17}_{-0.17}$, $C^{+0.25}_{-0.25}$, $C^{+0.335}_{-0.335}$, and $C^{+0.64}_{-0.64}$, and output states $C^{+0.425}_{+0.075}$, $C^{+0.52}_{-0.011}$, $C^{+0.6}_{-0.1}$, and $C^{+0.9}_{-0.41}$, respectively. From these results, the parameters $\gamma$, $\gamma'$, $\gamma''_-$, and $\gamma''_+$ are approximately related as $\gamma''_+\thickapprox \gamma' + \gamma$ and $\gamma''_-\thickapprox \gamma'-\gamma$, 
which is analogous to the observed in the interaction between the SB--WP in the state $C$ and a dispersive medium, expressed by Eq.~\eqref{eq:NewNotation}. We can then explicitly define the operation ``$\otimes$’’ between a SB signal in an arbitrary state and any of the linear systems studied, $T_D(\omega)$ or $T_A(\omega)$, as follows:
\begin{eqnarray}\label{eq:General_Interaction}
C^{\gamma_+}_{\gamma_-}\otimes H^{\gamma'_+}_{\gamma'_-} = C^{\gamma''_+}_{\gamma''_-}\\
\gamma''_{\pm} = \gamma'_{\pm} + \gamma_{\pm}\;\;\;\nonumber.
\end{eqnarray}
Based on Eq.~\eqref{e9}, the above results can be interpreted in terms of mathematical convolutions or deconvolutions. Indeed, the interaction $``\otimes"$ produces a new pair of \textit{chirpons}, each described by a Lorentzian function whose parameter $\gamma_{\pm}''$ is determined by the algebraic sum of the input ($\gamma_{\pm}$) and system ($\gamma_{\pm}'$) parameters. Since these parameters characterize the individual Lorentzian functions $L_{\gamma_{\pm}}(t)$ and $L_{\gamma'_{\pm}}(t)$ associated with the SB--WP and the linear system, respectively, each component locally satisfies either the convolution or deconvolution property (see Appendix A). This behavior corresponds to a local broadening or narrowing of the profile, which is strictly dictated by the relative signs of $\gamma_{\pm}$ and $\gamma_{\pm}'$. Hence, the \textit{chirpons} resulting from the interaction in Eq.~\eqref{eq:General_Interaction} can be generally expressed as
\begin{eqnarray}\label{eq:omega_conv}
\omega^{\pm}_{C''}(t,\gamma''_{\pm}) &=& \omega^{\pm}_{C''}(t,\gamma_{\pm} +\gamma'_{\pm}). %\\
%&=& \begin{cases}
%    \omega_{C}^{\pm}(t,\gamma)\circledast\omega_{C'}(t,\gamma') & \text{si } \gamma, \gamma' \gtrless 0, \\
%    \omega_{C}^{\pm}(t,\gamma)\circledast^{-1}\omega_{C'}(t,\gamma')  & \text{si} \gamma >0,\gamma' < 0, \\
%    \omega_{C}^{\pm}(t,\gamma')\circledast^{-1}\omega_{C'}(t,\gamma)  & \text{si } \gamma < 0, \gamma' > 0.
%\end{cases}
\end{eqnarray}

Having established the operational form of the \textit{chirpon} interaction, an arbitrary state $C^{\gamma_+}_{\gamma_-}$ can now be generated from $C$ by sequentially applying almost four linear systems,  $H^{+\gamma_1}_{-\gamma_1}\otimes H^{+\gamma_2}_{+\gamma_2}\otimes H^{-\gamma_3}_{-\gamma_3}\otimes H^{-\gamma_4}_{+\gamma_4}=H^{\gamma_+}_{\gamma_-}$, where $\gamma_{\pm}=\pm\gamma_1+\gamma_2-\gamma_3 \mp \gamma_4$. Physically, this corresponds to two dispersive media,  $H^{+\gamma_1}_{-\gamma_1}$ ($\Omega>0$) and $H^{-\gamma_4}_{+\gamma_4}$ ($\Omega < 0$), and two harmonic oscillators, $H^{+\gamma_2}_{+\gamma_2}$ ($\omega_r > \omega_0$) and $H^{-\gamma_3}_{-\gamma_3}$ ($\omega_r < \omega_0$).

Finally, it is worth noting that, in analogy with vortices in two-dimensional space~\cite{dennis2001topological}, a topological charge can be associated with each one-dimensional phase singularity through the function $\omega_C^{\pm}(t,\gamma)$: $q_C = \frac{1}{2\pi} \int_{-\infty}^{+\infty} \omega_C^{\pm}(t,\gamma)\, dt$,
%
%\begin{equation}\label{eq:qC}
%q_C = \frac{1}{2\pi} \int_{-\infty}^{+\infty} \omega_C^{\pm}(t,\gamma)\, dt \;,  
%\end{equation}
%
whose value is $+1/2$ ($-1/2$) for $\gamma > 0$ ($\gamma < 0$). 
Therefore, as a direct consequence of Eqs.~\eqref{eq:General_Interaction} and \eqref{eq:omega_conv}, the magnitude of this topological charge, $|q_C|$, remains conserved under linear interactions.

{\textit{Experimental}}.--- To experimentally confirm the emergence of \textit{chirpons}, we built an RLC circuit whose dynamics is equivalent to that of a damped harmonic oscillator in Eq.~\eqref{e16}.
This circuit has a quality factor $Q \approx 4.4$ and a resonance frequency $\nu_r \approx 8.7$ MHz. Two SB input signals were generated with a commercial arbitrary waveform generator, while the output was measured with an oscilloscope and mathematically processed to extract amplitudes and phases. In the first case, the SB signal was synthesized from a Gaussian signal with carrier frequency $\nu_0 = 6$ MHz (FWHM $\approx 1.18\mu$s) according to Eq.~\eqref{e1}, with $\alpha = 0.6$ and $\beta = 0.5$; in the second case, from a Gaussian signal with $\nu_0 = 12$ MHz (FWHM $\approx 0.59\mu$s), using the same values of synthesis parameters $\alpha$ and $\beta$. In Appendix B, we describe in detail the experimental setup and the signals generated.

\begin{figure}[h!]
\includegraphics[width=0.5\textwidth]{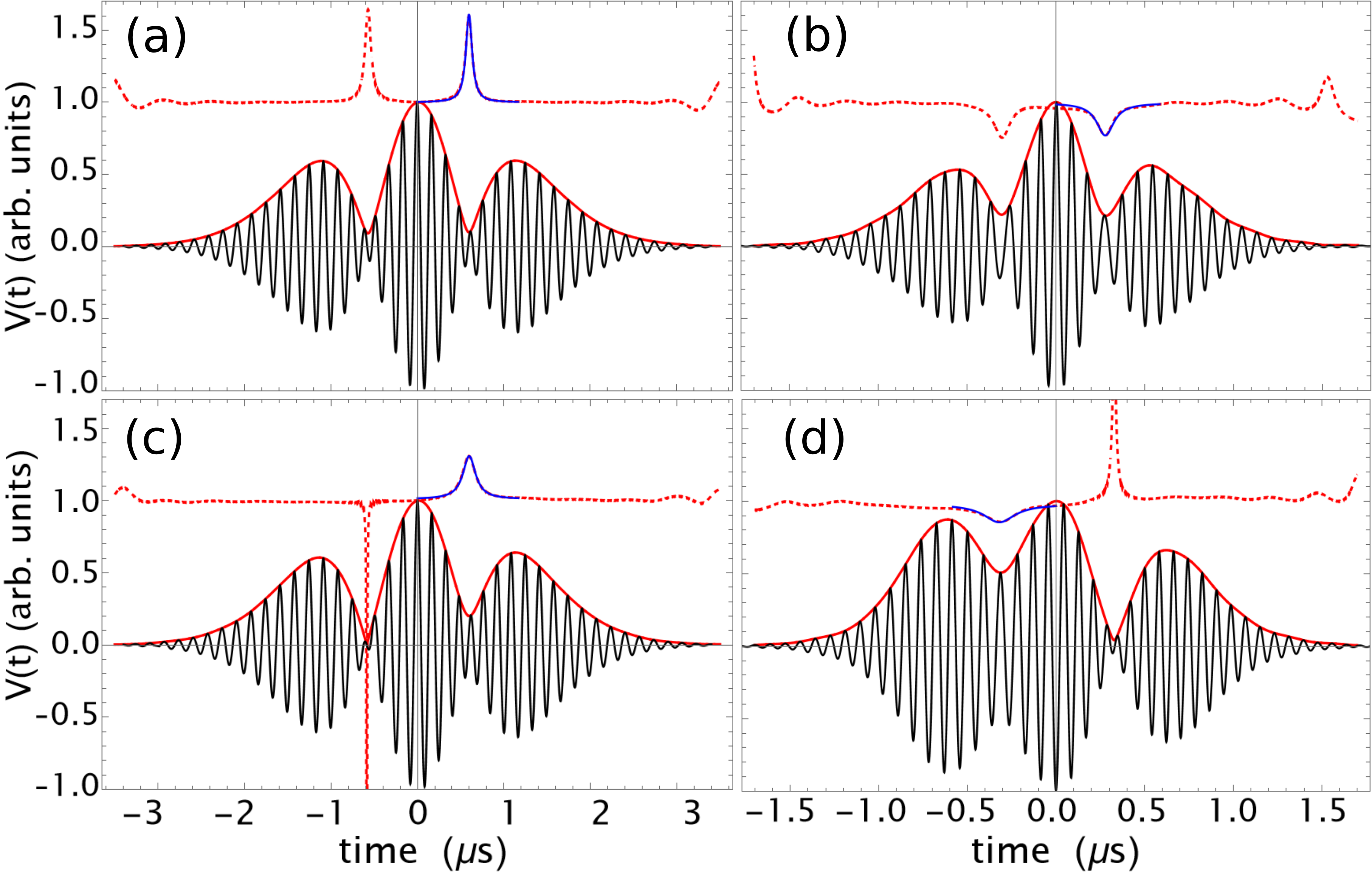}
\caption{Experimental results. Response of the RLC circuit to: an SB state $C$ with $\nu_0 = 6\;\mathrm{MHz} < \nu_r$ in (a) and $\nu_0 = 12\;\mathrm{MHz} > \nu_r$ in (b); (c) a state $C^{+\gamma^{(a)}}_{-\gamma^{(a)}}$, $\nu_0 = 6\;\mathrm{MHz} < \nu_r$, $\gamma^{(a)} > 0$; (d) a state $C^{+\gamma^{(b)}}_{-\gamma^{(b)}}$, $\nu_0 = 12\;\mathrm{MHz} > \nu_r$, $\gamma^{(b)} < 0$. The blue line at the top of each panel corresponds to the fit of the right \textit{chirpon} (dashed red line) in the output signal.} \label{fig5}
\end{figure}

In Figs.~\ref{fig5}(a) and \ref{fig5}(b), the response of the RLC circuit illustrates how the \textit{chirpons}—shown as the dashed red line in the upper part of the figures—arise from the singularities of the SB input signal $C$: for $\nu_0 = 6\;\mathrm{MHz} < \nu_r$ (Fig.~\ref{fig5}(a)), two \textit{up-chirpons} appear $\left(\omega_C^{\pm}(t_z,\gamma) > \omega_0\right)$, whereas for $\nu_0 = 12\;\mathrm{MHz} > \nu_r$ (Fig.~\ref{fig5}(b)) two \textit{down-chirpons} $\left(\omega_C^{\pm}(t_z,\gamma) < \omega_0\right)$ are observed.
The blue lines correspond to fits of the \textit{chirpon} shapes by means of the function $\omega_C(t,\gamma)$, with $\gamma_+^{(a)} = \gamma_-^{(a)} \equiv \gamma^{(a)} = 0.262T_0^{(a)}$ $\left(T_0^{(a)} = \tfrac{1}{6\;\mathrm{MHz}}\right)$ in Fig.~\ref{fig5}(a), and $\gamma_+^{(b)} = \gamma_-^{(b)} \equiv \gamma^{(b)} = -0.706T_0^{(b)}$ $\left(T_0^{(b)} = \frac{1}{12\;\mathrm{MHz}}\right)$ in Fig.~\ref{fig5}(b), confirming that this function captures the \textit{chirpons}. Also, in agreement with Eq.~\eqref{eq:NewNotation}, we obtain the operational notation for the \textit{chirpons} hosted by the RLC circuit: $H^{+\gamma^{(a)}}_{+\gamma^{(a)}}$ if $\nu_0=6$\;MHz, and $H^{+\gamma^{(b)}}_{+\gamma^{(b)}}$ if $\nu_0=12$\;MHz.
 
We next investigate the system response under alternative input signals to test the more general interaction described by Eq.~\eqref{eq:General_Interaction}.
Using the arbitrary waveform generator, we synthesized the states
$C^{+\gamma^{(a)}}_{-\gamma^{(a)}}$ with $\nu_0=6$\;MHz and $C^{-\gamma^{(b)}}_{+\gamma^{(b)}}$ with $\nu_0=12$\;MHz, which were subsequently used to drive the same RLC circuit studied above. Figures~\ref{fig5}(c) and \ref{fig5}(d) show the corresponding output signals, $C^{\gamma^{(c)}_+}_{\gamma^{(c)}_-}$ and $C^{\gamma^{(d)}_+}_{\gamma^{(d)}_-}$, respectively. As in the previous case, the $\gamma$ parameters were extracted from the fitting curve: $\gamma^{(c)}_+= 0.532\,T_0^{(c)}$ $\left(T_0^{(c)} = \frac{1}{6\,\mathrm{MHz}}\right)$ and $\gamma_-^{(d)} = -1.2816\,T_0^{(d)}$ $\left(T_0^{(d)} = \frac{1}{12\,\mathrm{MHz}}\right)$, thus verifying the relations $C^{+\gamma^{(a)}}_{-\gamma^{(a)}}\otimes H^{+\gamma^{(a)}}_{+\gamma^{(a)}} = C^{\gamma^{(c)}_+}_{\gamma^{(c)}_-}$,
$\gamma_+^{(c)} \thickapprox 2\gamma^{(a)}$, $\gamma_-^{(c)} \thickapprox 0$, and, with less accuracy, $C^{-\gamma^{(b)}}_{+\gamma^{(b)}}\otimes H^{+\gamma^{(b)}}_{+\gamma^{(b)}} = C^{\gamma^{(d)}_+}_{\gamma^{(d)}_-}$, $\gamma_+^{(d)} \thickapprox 0$, $\gamma_-^{(d)} \thickapprox 2\gamma^{(b)}$. 

{\textit{Outlook}}.---Although often overlooked, phase singularities---points where the field vanishes and the phase changes abruptly---play a fundamental role in wave dynamics. In particular, one-dimensional phase singularities can be identified in diverse WPs, such as complex-spectrum few-cycle pulses~\cite{viotti2022multi,manzoni2015coherent}, derived Gaussian signals~\cite{sol2022meta,slavik2006ultrafast}, and as sources of superoscillations~\cite{yuan2019plasmonics,berry2008natural}, yet their dynamical consequences have not been fully explored. 

Here, we have shown that these singularities give rise to localized oscillations when the WP interacts with a linear system. Depending on the system’s response, these oscillations---which we termed \textit{chirpons}---can occur beyond the limits of the WP's Fourier spectrum, and are therefore not captured by standard spectral methods. 
By studying the interaction of SB--WPs with a damped harmonic oscillator, we have identified the distinctive dynamics of \textit{chirpons} and provided a mathematical characterization through a real parameter $(\gamma)$, offering a first understanding of their behavior and extending the scope of spectral analysis. In particular, we demonstrated that the peak instantaneous frequencies of \textit{chirpons}, when interacting with resonant systems, are governed by the system resonance instead of the carrier frequency of the input signal, highlighting a behavior that cannot be captured by a purely Fourier spectrum-based description.

Moreover, from the function describing each \textit{chirpon}, a topological charge can be assigned to the singularity that acts as its source. The absolute value of this charge is conserved under linear interactions meaning that the phase structure around the singularity is preserved: a new \textit{chirpon} with a different value of $\gamma$ emerges, so linear systems can be regarded as operators acting on the \textit{set of chirpons}. From an operational point of view, this formulation, confirmed by the experimental results, enables the design and manipulation of \textit{chirpons} at prescribed frequencies, showing that their properties can be controlled in a predictable manner, consistent with information obtained from numerical fittings. 

Beyond a fundamental interest, this establishes \textit{chirpons} as physically observable, singularity-triggered dynamical events, rather than as mere abstract features of the input field. Even more, the presence of these oscillations also suggests potential applications: 
their association with abrupt variations in the WP could make them sensitive probes for sensing purposes. 
In fact, related approaches have been explored using time-domain superoscillatory fields~\cite{peng2025super}, subwavelength resolution in radar~\cite{jordan2023fundamental} and imaging~\cite{grant2025localization}. In this regard, the present framework offers an alternative and practical way to construct such probes and to anticipate the response of a given linear system, pointing to the potential of \textit{chirpons} as a tool for exploring and characterizing linear dynamics.

\section*{Acknowledgements}
This work was partially supported by CONICET, CNEA, UNCuyo (Projects SIIP 2022-2024 06/C034-T1 and SIIP 2022-2024 06/C033-T1), ANPCyT (Projects PICT-2020-SERIE A-823 and PICT-2021-GRF-TI-00742), Red Federal
de Alto Impacto FFFLASH MinCyT, Fondo Semilla para Emprendimientos Tecnológicos 2024 (CNEA - Fundación Balseiro) and Fundación Sadosky.

\appendix

\section{Appendix A. Convolution and deconvolution}\label{app:conv}

If $L_{\gamma}(t)=\frac{1}{\pi}  \frac{\gamma}{t^2 + \gamma^2}$, then for $\omega \geq 0$ its Fourier transform is 
\begin{eqnarray}    
\mathcal{F}\left\{L_{\gamma}(t)\right\}=\int^{\infty}_{-\infty}L_{\gamma}(t)e^{-i t\omega}dt = \mathrm{sgn}(\gamma) \; e^{-|\gamma|\omega},
\end{eqnarray}
where $\operatorname{sgn}(\gamma) = \frac{\gamma}{\vert{}\gamma\vert{}}$. Hence, for $\gamma_1,\;\gamma_2\neq 0$, the Fourier transform of the convolution between $L_{\gamma_1}(t)$ and $L_{\gamma_2}(t)$ satisfies 
\begin{eqnarray}
\mathcal{F}\left\{L_{\gamma_1}(t)\circledast L_{\gamma_2}(t)\right\}&=&\mathrm{sgn}(\gamma_1\gamma_2)\; e^{-|\gamma_1|\omega}e^{-|\gamma_2|\omega}\nonumber\\
&=&\mathrm{sgn}(\gamma_1\gamma_2)\; e^{-(|\gamma_1|+|\gamma_2|)\omega}, 
\end{eqnarray}
which implies that 
\begin{eqnarray}
L_{\gamma_1}(t) \circledast L_{\gamma_2}(t)&=&\mathcal{F}^{-1}\left\{\operatorname{sgn}(\gamma_1 \gamma_2)\;e^{-(|\gamma_1|+|\gamma_2|)\;\omega}\right\}\nonumber\\
&=& \frac{\operatorname{sgn}(\gamma_1 \gamma_2)}{\pi} \frac{\vert{}\gamma_1\vert{}+\vert{}\gamma_2\vert{}}{t^2 + (\vert{}\gamma_1\vert{}+\vert{}\gamma_2\vert{})^2}.
\end{eqnarray}

On the other hand, if a Lorentzian function $L_{\Gamma}(t)$ is obtained from the convolution of two Lorentzian functions $L_{\gamma_1}(t)$ and $L_{\gamma_2}(t)$, such that $L_{\Gamma}(t) = L_{\gamma_1}(t) \circledast L_{\gamma_2}(t)$, then the deconvolution process allows recovering one of the components (e.g., $L_{\gamma_1}(t)$) by inverting the operation. In the frequency domain ($\omega \geq 0$), the deconvolution operator $\circledast^{-1}$ satisfies:
\begin{eqnarray}
\mathcal{F}\left\{L_{\Gamma}(t) \circledast^{-1} L_{\gamma_2}(t)\right\} &=& \frac{\mathcal{F}\left\{L_{\Gamma}(t)\right\}}{\mathcal{F}\left\{L_{\gamma_2}(t)\right\}} 
= \frac{\operatorname{sgn}(\Gamma) e^{-\vert{}\Gamma\vert{}\omega}}{\operatorname{sgn}(\gamma_2) e^{-\vert{}\gamma_2\vert{}\omega}} \nonumber\\
&=& \operatorname{sgn}(\Gamma\;\gamma_2) \, e^ {- (\vert{}\Gamma\vert{} - \vert{}\gamma_2\vert{})\omega}.
\end{eqnarray}
Provided that $\vert{}\Gamma\vert{} > \vert{}\gamma_2\vert{}$, thus the inverse Fourier transform yields:
\begin{eqnarray}
L_{\Gamma}(t) \circledast^{-1} L_{\gamma_2}(t) &=& \mathcal{F}^{-1}\left\{\operatorname{sgn}(\Gamma \;\gamma_2) \, e^ {- (\vert{}\Gamma\vert{} - \vert{}\gamma_2\vert{})\omega}\right\} \nonumber\\
&=& \frac{\operatorname{sgn}(\Gamma \;\gamma_2)}{\pi} \frac{\vert{}\Gamma\vert{} - \vert{}\gamma_2\vert{}}{t^2 + (\vert{}\Gamma\vert{} - \vert{}\gamma_2\vert{})^2} \nonumber\\
&=& \operatorname{sgn}(\Gamma \;\gamma_2) \, L_{\vert{}\Gamma\vert{} - \vert{}\gamma_2\vert{}}(t)\,.
\end{eqnarray}

It is worth noting that, based on these properties, the interaction between \textit{chirpons} (Eq.~\eqref{eq:General_Interaction}), which yields the relation in Eq.~\eqref{eq:omega_conv}, represents a convolution or deconvolution acting on the geometrical local profile ($\gamma$ parameter) rather than on the temporal coordinate, since singularities appear at the exact same instant as the input pulse.

\section{Appendix B. Experimental setup} \label{app:setup}

\begin{figure}[h!]
\includegraphics[width=0.45\textwidth]{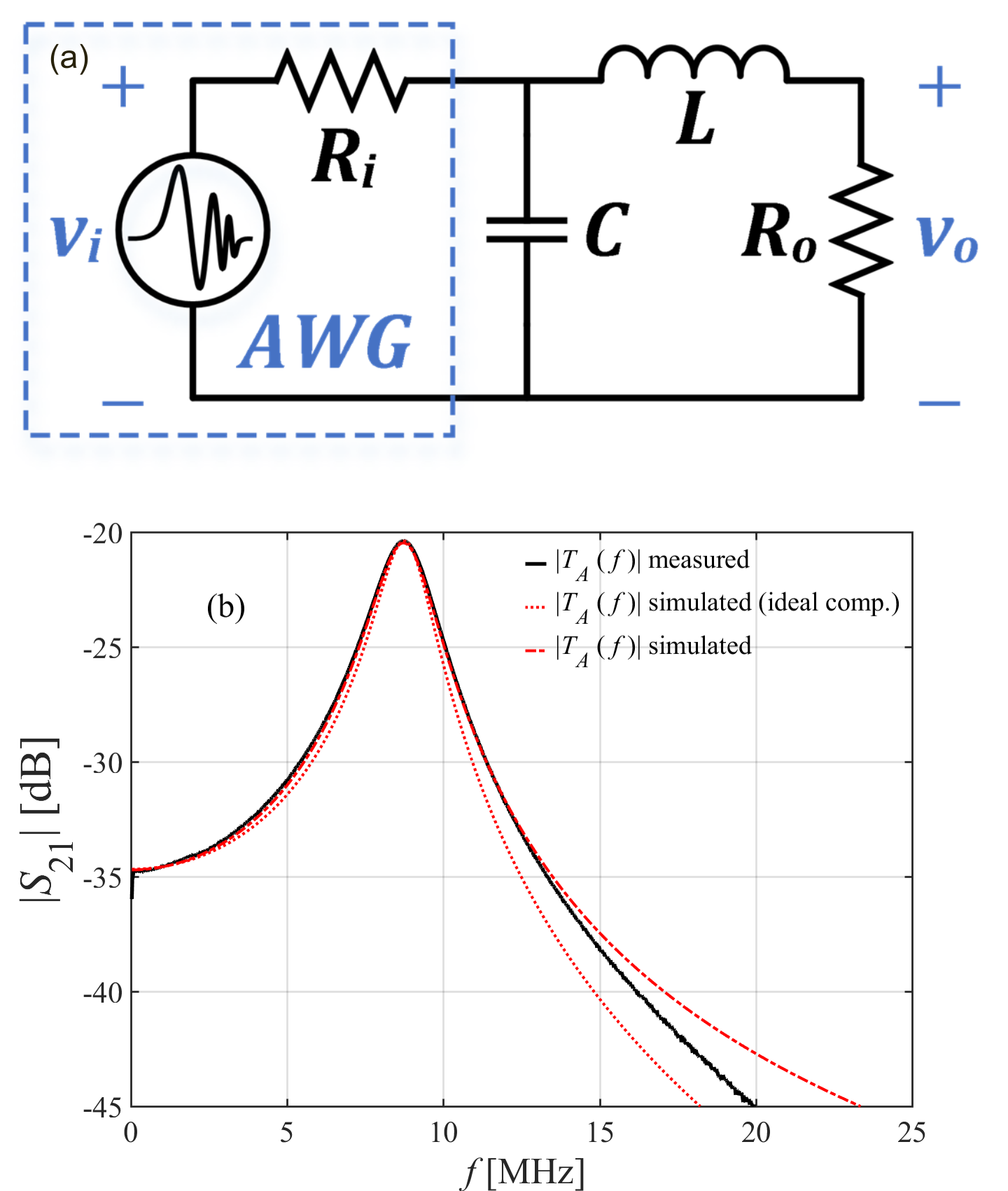}
\caption {(a) Schematic diagram of the experimental setup used to measure the interaction of the \textit{chirpons} with a damped harmonic oscillator. AWS: Arbitrary Wave Form Generator. (b) Transfer function of the damped harmonic oscillator: measured in solid black line, simulated with ideal components in dotted red line, simulated including non ideal models for the inductor and the resistor $R_0$ in dashed red line.
}\label{fig1A}
\end{figure}

Figure \ref{fig1A}(a) shows the experimental setup used to obtain the response of a linear system---a damped harmonic oscillator based on a parallel RLC circuit---when it interacts with the \textit{chirpons}. Resistor $R_0$ is introduced to indirectly measure the inductor’s current. It can be easily shown that the differential equation that governs the circuit behavior is:
\begin{eqnarray}\label{eq:A1}
\frac{R_0}{R_i}\frac{1}{LC}v_i(t)&=&\frac{d^2 v_0(t)}{dt^2}\\ 
&+&\left(\frac{R_0}{L}+\frac{1}{R_iC}\right)\frac{d v_0(t)}{dt}
+\left(1+\frac{R_0}{R_i}\right)\frac{1}{LC} v_0(t).\nonumber
\end{eqnarray} 
Equation~\eqref{eq:A1} is rewritten in terms of the resonance frequency $\omega_r$ and the quality factor $Q$: 
\begin{equation}\label{eq:A2}
\left(\omega_r^2-\omega_n^2\right) v_i(t)\;=\frac{d^2 v_0(t)}{dt^2}+\frac{\omega_r}{Q}\frac{d v_0(t)}{dt}+\omega_r^2\; v_0(t),
\end{equation} 
where $\omega_r^2=(1+\frac{R_0}{R_i})~\omega_n^2$ with $\omega_n^2=\frac{1}{LC}$, and $Q^{-1} = Q_s^{-1}+Q_p^{-1}$ with $Q_s=\omega_r\frac{L}{R_0}$, $Q_p=\omega_rR_iC$. Since Eq.~\eqref{eq:A2} has the same form as Eq.~\eqref{e16}, the transfer function of the corresponding circuit is 
\begin{eqnarray}
T_A(\omega)=\frac{\omega_r^2-\omega_n^2}{\omega_r^2-\omega^2+i\omega_r\omega/Q}
\end{eqnarray}
which is, except for a constant factor, exactly the same as that presented in the main manuscript. 

Figure \ref{fig1A}(b) shows the transmission $s$-parameter, $S_{21}(f)$, of the transfer function $T_A(f)$ ($f=\omega/2\pi$), measured with a Vector Network Analyzer (solid black line). The simulation, based on the ideal model of each component with nominal values $C=3.28$nF, $L=100$nH, $R_i=50\Omega$, and $R_0=0.47\Omega$, which yields the nominal values $\nu_r=8.83$MHz and $Q=5.1$, is shown as a red dotted line. 
Although this curve deviates from the measured response, it clearly reproduces the overall behavior.
\begin{figure}[h!]
\includegraphics[width=0.5\textwidth]{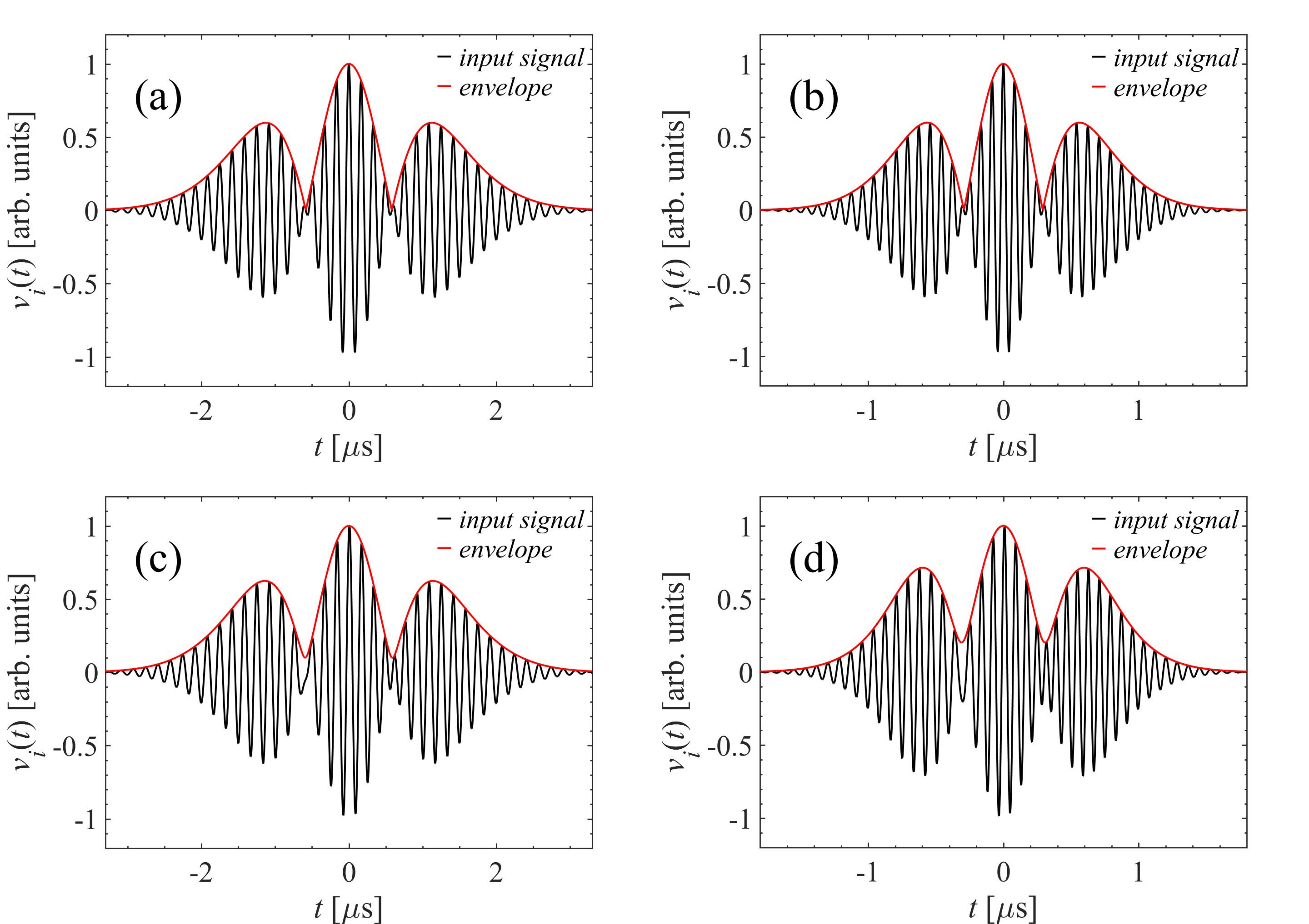}
\caption {Input signals to the resonant circuit of Fig.~\ref{fig1A}(a) generated with an AWG. Panels (a) and (b) correspond to two SB signals with $\beta=0.5$ and $\alpha=0.6$, centered at $\nu_0=6$ MHz and $\nu_0=12$ MHz, respectively. Panels (c) and (d) show the generated signals corresponding to \textit{chirpons} in the states $C^{+\gamma^{(a)}}_{-\gamma^{(a)}}$ and $C^{-\gamma^{(b)}}_{+\gamma^{(b)}}$, respectively
}
 \label{fig2A}
\end{figure}
If the component model is then modified to include the most significant non-idealities---specifically, the finite $Q$, the tolerance of the inductor, and the stray inductance of the resistor $R_0$---the simulated response almost perfectly fits the measured curve in the frequency range of interest (red dashed line).
Therefore, the resonant frequency and the quality factor are slightly reduced compared to the ideal case, matching the measured values $\nu_r=8.73$ MHz and $Q=4.4$. Extending the agreement between simulation and experiment to higher frequencies (e.g., above 15 MHz) requires a more accurate capacitor model, with a stray series inductance included to account for its self-resonant frequency.

The input signals synthesized with an arbitrary wave generator (AWG), which  drive the RLC circuit of Fig.~\ref{fig1A}(a) to produce the experimental results presented in Fig.~\ref{fig5} of the main manuscript, are shown in Fig.~\ref{fig2A}. Panels (a) and (b) correspond to two SB signals with $\beta=0.5$ and $\alpha=0.6$, centered at $\nu_0=6$ MHz and $\nu_0=12$ MHz, respectively. Panels (c) and (d) show the generated signals corresponding to \textit{chirpons} in the states $C^{+\gamma^{(a)}}_{-\gamma^{(a)}}$ and $C^{-\gamma^{(b)}}_{+\gamma^{(b)}}$, respectively. For generating these states, the dispersive parameter $\Omega$ ($T_D(\omega) = e^{-i\Omega(\omega - \omega_0)^2}$) are $\Omega= 0.68[T_0^{(a)}]^2$ for $C^{+\gamma^{(a)}}_{-\gamma^{(a)}}$ and $\Omega=2.20[T_0^{(b)}]^2$ for $C^{-\gamma^{(b)}}_{+\gamma^{(b)}}$.

\bibliography{main}

\end{document}